\documentclass[acmlarge]{acmart}

\usepackage{caption}
\usepackage{subcaption}
\usepackage[vskip=1em,font=itshape,leftmargin=2em,rightmargin=2em]{quoting}
\usepackage{amsmath}
\usepackage{makecell}

\newcommand{\pquote}[1]{{\textit{``#1''}}}
\newcommand{\dquote}[1]{{``#1''}}

\newcommand{\added}[1]{\textcolor{black}{#1}}
\newcommand{\deleted}[1]{\textcolor{lightgray}{}}

\begin{document}

\title[Trustless Autonomy]{Trustless Autonomy: Understanding Motivations, Benefits and Governance Dilemmas in Self-Sovereign Decentralized AI Agents}



\author{Botao Amber Hu}
\orcid{0000-0002-4504-0941}
\affiliation{%
  \institution{Reality Design Lab}
  \city{New York City}
  \country{USA}
  }
\email{amber@reality.design}

\author{Yuhan Liu}
\orcid{0000-0001-6852-6218}
\affiliation{%
  \institution{Princeton University}
  \city{Princeton}
  \state{New Jersey}
  \country{USA}
  }
\email{yuhanl@princeton.edu}

\author{Helena Rong}\authornote{Corresponding author}
\orcid{0000-0003-1626-7968}
\affiliation{%
  \institution{New York University Shanghai}
  \city{Shanghai}
  \country{China}
  }
\email{hr2703@nyu.edu}


\begin{abstract}
The recent trend of self-sovereign Decentralized AI Agents (DeAgents) combines Large Language Model (LLM)-based AI agents with decentralization technologies such as blockchain smart contracts and trusted execution environments (TEEs). These tamper-resistant trustless substrates allow agents to achieve self-sovereignty through ownership of cryptowallet private keys and control of digital assets and social media accounts. DeAgents eliminate centralized control and reduce human intervention, addressing key trust concerns inherent in centralized AI systems. \added{This contributes to social computing by enabling new human cooperative paradigm "intelligence as commons."} However, given ongoing challenges in LLM reliability such as hallucinations, this creates paradoxical tension between trustlessness and unreliable autonomy. This study addresses this empirical research gap through interviews with DeAgents stakeholders—experts, founders, and developers—to examine their motivations, benefits, and governance dilemmas. The findings will guide future DeAgents system and protocol design and inform discussions about governance in sociotechnical AI systems in the future agentic web. \end{abstract}





\maketitle

\section{Introduction}


The recent integration of AI agents with decentralization technologies \cite{Karim2025AI,Bhumichai2024Convergence}—making agents self-sovereign beyond mere autonomy—has become a surging phenomenon. This trend is most commonly known as Decentralized AI Agents (DeAgents) \cite{Wang2025SoK}, meaning the deployment of Large Language Model (LLM)-based AI agents \cite{Wang2024survey} on trustless \cite{Harz2019Scalability} (no need to trust third party) tamper-resistant computational substrates, including blockchain smart contracts \cite{Buterin2014}, decentralized physical infrastructures (DePIN) \cite{Ballandies2023Taxonomy}, and trusted execution environments (TEEs) \cite{Sabt2015Trusted}. Self-sovereignty means that once launched, an agent holds its own cryptographic private keys and makes autonomous decisions without human intervention. These agents can manage their cryptocurrency wallets, transfer digital assets, interact with decentralized finance (DeFi) protocols, issue tokens for fundraising, and maintain social media accounts to build influence and incentivize followers through both narrative and financial means—all "free from human control" \cite{Hu2025Decentralized}. The phenomenon debuted with \emph{Truth Terminal} \cite{Ante2025Transforming} on June 17, 2024—an autonomous chatbot that independently manages its Twitter account and its own crypto wallet, becoming the first crypto millionaire. The technology was subsequently democratized through the open-source decentralized AI framework \emph{ElizaOS} \cite{Walters2025Eliza} and further popularized by \emph{Virtuals Protocol}, a low-code toolkit for building tokenized AI agents. Despite the speculative nature of the crypto market \cite{Kukacka2023Fundamental, Baur2018Bitcoin}, the total market capitalization of all DeAgents reached \$10 billion by December 2024 \cite{Sentient2024}.
 

Deploying DeAgents on trustless secure enclaves, such as smart contracts and TEEs, inherently creates paradoxical tensions around governance risks due to diminished human oversight. Proponents envision that self-sovereign DeAgents can help moderate decentralized autonomous organizations (DAO) \cite{Hinkes2024Artificial}, manage financial portfolios \cite{Ante2024Autonomous}, and etc., aiming to mitigate human error and centralized corruption \cite{Trequattrini2024contribution, Odilla2023Bots}. For example, \emph{ai16z} is the first AI venture capital fund that is fully managed by DeAgent with recommendations from DAO members \cite{ai16z}. However, while trustless infrastructure can guarantee tamper-resistant autonomous execution \cite{Jauernig2020Trusted}, the AI agent itself may not be inherently trustworthy \cite{Liu2024Trustworthy}. Large language models are prone to errors \cite{NEURIPS2024_6aebba00}, biases \cite{Rettenberger2025Assessing}, deceptions \cite{Hubinger2024Sleeper}, hypersensitive to nudges \cite{Cherep2025LLM}, hallucinations \cite{Huang2025Survey, Xu2025Hallucination}, and misalignment \cite{Shen2023Large}. The additional challenge of "locking" AI agents in secure enclaves to prevent external tampering arguably amplifies potential harms if the agents' decisions are flawed or hacked \cite{Patlan2025Real}—especially without a kill switch \cite{Seneviratne2024Feasibility}. This tension between trustlessness and autonomy fundamentally challenges the governability \cite{Hu2025Decentralized}, accountability \cite{Novelli2024Accountability} and responsibility \cite{Dignum2019Responsible} of sociotechnical AI systems \cite{Kudina2024sociotechnical,Dobbe2024Sociotechnical,Feng2025Sociotechnicala}.


Despite inherent paradoxical governance tensions, the reality is that thousands of DeAgents are increasingly being deployed for various purposes so far \cite{Sentient2024}. As this trend continues with significant societal and financial impacts, there remains a significant gap in empirical research about why stakeholders prefer DeAgents over traditional AI agents. To address this gap, we conducted a qualitative study through semi-structured interviews with DeAgents stakeholders to understand why developers, founders, and experts embrace DeAgents—exploring their motivations, expectations, and perceived benefits while examining the practical challenges and governance dilemmas they face. Specifically, we examined:

\begin{itemize}
    \item \textbf{RQ1} (Motivations and Expectations): What motivates developers and communities to deploy DeAgents, and what are their expectations?
    \item \added{\textbf{RQ2} (Emerging Social Dynamics): What roles do DeAgents come to play in human cooperation and social computing and how do people perceive them?}
    \item \textbf{RQ3} (Challenges and Governance Implications): What challenges arise in the real-world deployment of DeAgents, and how do stakeholders navigate governance, oversight, and accountability within a \emph{trustless autonomous system}?
\end{itemize}

Grounded in thematic analysis, our findings illuminate a spectrum of stakeholder perspectives—from enthusiastic proponents who see DeAgents as a breakthrough for autonomy and censorship-resistance, to cautious adopters and critics concerned about unpredictable behavior and accountability voids. We reveal paradoxical tensions in the quest to fuse AI autonomy with trustless infrastructure. Our contributions are threefold:

\begin{itemize}
\item 
We presented \textbf{empirical findings} showing that the push for self-sovereign AI agents on trustless substrates is driven by sociotechnical / political factors like decentralization, privacy concerns, censorship resistance, and community ownership—even as AI reliability issues remain unresolved. \added{Our study further identified DeAgents’ social impacts, including the evolving role they play, where the attention and trust they derived from. We summarized the difference between traditional AI agents and DeAgents.}
\item We identified the \textbf{paradoxical governance dilemma} that arises when potentially unreliable AI agents operate within trustless infrastructure, revealing the complex tensions between trusting humans versus trusting AI in delegating agency to autonomous systems. \added{Moreover, we found that the decentralized nature of DeAgents is the foundation to function as \dquote{commons} for social good, as it is manipulation-free. }
\item We proposed \textbf{design and governance implications} that \added{potentially address governance dilemma and} emphasize thoughtful protocol design for agentic web of trust, with flexible fail-safes serving as essential bridges between trustlessness and genuine trustworthiness.
\end{itemize}


\section{Background and Related Works}

\subsection{Terminology used in this paper}

This paper discusses Decentralized AI, which involves technical terminology that may carry different meanings across contexts and may be unfamiliar to readers who are new to blockchain and cryptocurrency. Therefore, we provide the following definitions based on prior work.

\paragraph{Trust} The confidence that another agent or system will behave reliably and act in accordance with one's interests or expectations \cite{Werbach2016Trustless}.

\paragraph{Trustworthiness} The quality indicating the likelihood that an agent or system will consistently behave as expected, making it deserving of trust \cite{Duran2025Trust}.

\paragraph{Trustlessness} The property of an agent or system that removes the need to place belief in any single third party; instead, cryptography and consensus provide objective verifiability, although this often merely relocates trust—to code, hardware, or governance processes \cite{Werbach2016Trustless}. 

\paragraph{Permissionless} The property of a system, network, or protocol that does not require authorization or approval from a central authority for participation. Anyone can join, interact, validate, and contribute to the network without needing special permission \cite{Nabben2022Permissionlessness}.

\paragraph{Autonomy} The capacity of an agent to perceive, decide, and act toward its goals without continuous human oversight, also known as "self-governance." However, higher powers may still exist that can assume control over the agent \cite{Hui2024Machine, feng2025levels}.

\paragraph{Sovereignty} The capability of an agent to independently control, manage, and govern its own resources, behaviors, and interactions, operating in an environment where no higher power can override its decisions \cite{Hui2024Machine}.

\paragraph{Decentralized AI} An architectural paradigm in which the AI’s computation, data, and governance are distributed across a network of independent nodes that coordinate through consensus or cryptographic protocols, eliminating reliance on any single authority \cite{Singh2024Perspective, Wang2025SoK}.

\paragraph{Decentralized AI Agents} Autonomous software agents whose model, data, and execution are distributed across a network of independent nodes that coordinate through cryptographic protocols, thereby eliminating single-point authority while enabling verifiable, fault-tolerant, self-sovereign \cite{Hu2025Decentralized}.

\paragraph{Trustless Agent} \added{Agents that incorporate a trust layer enabling participants to discover, select, and engage with agents across organizational boundaries without requiring pre-existing trust } \cite{ethereumERC8004Trustless}. 

\paragraph{Decentralized Autonomous Organization (DAO)} \added{DAO is an organization managed in whole or in part by decentralized computer programs, with voting and finances handled through a decentralized ledger technology like a blockchain.
}


\subsection{LLM-based Agents}









Recent advancements in Large Language Models have given rise to autonomous agents \cite{Wang2024survey, Xi2023Rise} capable of performing various real-world tasks \cite{Xu2024TheAgentCompany}, such as automatic social media marketing \cite{Deshmukh2025Harnessing}, scientific discovery \cite{Ren2025Scientific,Gridach2025Agentic} and financial trading \cite{Ding2024Large}.  
These LLM-based agents typically share a common architectural framework \cite{Luo2025Large} that includes profile definition, memory mechanisms \cite{Zhang2024Survey}, planning capabilities \cite{Huang2024Understanding,He2025PlanThenExecute}, action execution via context protocol and external tools \cite{Hou2025Model}, inter-agent collaboration \cite{Mou2024Individual}, and agent evolution \cite{Dai2024Artificial}.
Open-source agentic frameworks such as LangChain and AutoGen \cite{Wu2023AutoGen} make it trivial to wrap an LLM in an agent loop (plan → reason → act → observe). 

Traditionally, most LLM-based autonomous agents still run on a central cloud host. Every step in the agent loop triggers an LLM API call, resulting in ongoing costs for inference services (OpenAI, Anthropic, etc.) borne by the agent owners, who retain the ultimate control to terminate the agent’s activity at any moment. Centralization also gives the LLM API platform operator a kill-switch: an account ban or API deprecation can halt the agent instantly. However, the emergence of fully open-weight  foundational models, such as Llama \cite{Grattafiori2024Llama}, DeepSeek \cite{DeepSeek-AI2025DeepSeekR1}, and Qwen \cite{Qwen2025Qwen25}, has facilitated a shift toward deployment in edge computing environments \cite{Deng2020Edge} and private clouds. This decentralization potentially reduces operational costs, enhances privacy, and increases agent autonomy and control over data and computing resources.

Although LLM-based agents have already been deployed with commerical usage in the real world, LLM-based agents remain subject to significant trustworthiness concerns and reliability challenges, due to LLM inherent issues such as errors \cite{NEURIPS2024_6aebba00}, biases \cite{Rettenberger2025Assessing}, hypersensitive to nudges \cite{Cherep2025LLM}, deceptions \cite{Hubinger2024Sleeper}, hallucinations \cite{Huang2025Survey, Xu2025Hallucination}, and misalignment \cite{Shen2023Large}, where the agent generates incorrect, not aligned with human value content or even harmful actions. While human-feedback reinforcement learning (RLHF) methods have been implemented to solve this alignment problem to bias reduction and harm minimization, the process provides no inherent guarantee of alignment with ethical values or desired behavioral norms \cite{Barnhart2025Aligning}. Researchers demonstrate “reverse alignment” attacks: with a few hours of fine-tuning, an open-weight model’s safety guardrails can be removed entirely \cite{Yi2024Vulnerability}. The proliferation of open models, while offering greater flexibility for customized model fine-tuning, intensifies these alignment challenges since anyone can fine-tune models in potentially harmful ways—even without awareness—operating outside the oversight of model companies and traditional governance like law \cite{mo2023trustworthy, Hu2025Decentralized}.

From an ontological perspective, current LLM-based agents operate on fixed pre-trained models. Their identity and behavior emerge from three elements: fixed profile definition prompts, memory constructs \cite{Xu2025AMEM}, and interactions with their environment. While these agents employ memory enhancement techniques like Retrieval-Augmented Generation (RAG) \cite{Singh2025Agentic} to access past interactions and relevant context, their memory systems remain vulnerable to decay, injection, and poisoning attacks \cite{Patlan2025Real}. Unlike biological brains with neural plasticity, these agents cannot truly learn from experience—negative feedback is merely stored in memory constructs while the underlying model remains unchanged. This means agents cannot fundamentally adjust their behavior based on past experiences through external memory alone. Yet emerging paradigms like Liquid AI \cite{Hasani2020Liquid} propose lifelong learning models \cite{chen2018lifelong} that enable continuous adaptation and knowledge expansion. These evolving learning capabilities lay the groundwork for developing more adaptive and contextually aware agent behaviors.

LLM-based agents achieve interoperability by interacting with other agents and the world through protocols that integrate tools, share contextual data, and coordinate tasks across heterogeneous systems in a decentralized and permissionless way \cite{Ehtesham2025survey}. Four emerging agent communication protocols address interoperability in distinct deployment contexts: Model Context Protocol (MCP) \cite{Hou2025Model}, Agent Communication Protocol (ACP), Agent-to-Agent Protocol (A2A), and Agent Network Protocol (ANP). MCP provides a JSON-RPC client-server interface for secure tool invocation and typed data exchange. ACP enables REST-native messaging through multi-part messages and asynchronous streaming for multimodal agent responses. A2A facilitates enterprise-scale workflows through peer-to-peer task outsourcing using capability-based Agent Cards. ANP enables open-network agent discovery and secure collaboration using decentralized identifiers (DIDs) and JSON-LD graphs. These protocols differ across multiple dimensions, including interaction modes, discovery mechanisms, communication patterns, and security models—signaling the first steps toward a fully interoperable agentic web of billions of agents on the horizon.

\subsection{Trustless Computational Substrates}







Trustless computation substrates are computing infrastructures that enable programs or agents to run without requiring trust in any central authority. These substrates are permissionless—meaning anyone can use them without third-party approval—and tamper-resistant, protecting against unauthorized modifications. The three main types of trustless substrates are blockchain smart contracts, Trusted Execution Environments (TEEs), and Decentralized Physical Infrastructure Networks (DePIN).

Blockchain smart contracts \cite{Buterin2014} functions like a global, Turing-complete computer: any logic that can be written in code can, in principle, run atop its distributed network. Because every node independently verifies each instruction against a shared consensus protocol, the system delivers “trustless” computation—developers and users do not have to rely on a single operator or intermediary to guarantee correctness. Instead, the substrate of cryptographic proofs, decentralized validation, and immutable state ensures that programs execute exactly as written, producing tamper-evident results that anyone can audit. 

Trusted Execution Environments (TEEs) further reinforce this trustless paradigm by enabling secure, verifiable computation \cite{Li2023Survey}. A TEE is a hardware-based enclave that runs code isolated from the host system's OS and even the hypervisor \cite{Munoz2023survey}. Leading TEE technologies—including Intel SGX, ARM TrustZone, and AMD SEV—seal off sections of CPU memory from the outside world. The Nvidia Blackwell architecture also supports GPU TEEs for confidential machine learning \cite{Chaudhuri2025Secured}. TEEs enable privacy-preserving "Confidential Computation," preventing cloud services from accessing sensitive data within the TEE \cite{Lee2024PrivacyPreserving, Narra2019PrivacyPreserving}. Only authorized code can access data inside a TEE, and neither the machine's administrator nor operating system can tamper with or inspect the enclave's execution. When an LLM-based agent stores its private keys and policy weights inside a TEE, neither cloud administrators nor malicious node operators can "reach into the jar." This grants the agent a degree of sovereignty beyond traditional cloud sandboxes while remaining auditable from the chain. TEEs also provide verifiable computation through remote attestation—the enclave produces proof that its running bytecode matches an expected hash, enabling trustworthy results even on untrusted infrastructure. For example, an AI agent's logic could run inside a TEE on a volunteer's node, with the blockchain network and users receiving cryptographic assurance that the agent's code executed correctly and remained unaltered by the node operator. In essence, TEEs provide a trusted foundation that strengthens the trustless design: they guarantee integrity and confidentiality of computation, which is especially valuable for AI tasks requiring private data handling or off-chain processing \cite{Li2022SoK, Paju2023SoK}.

Decentralized Physical Infrastructure Networks (DePIN) \citep{Lin2025Decentralized,Fan2024New}—is the umbrella term for blockchain projects that crowd-source real-world hardware and reward participants with tokens, turning everything from wireless hotspots, solar panels to GPUs, storage drives and mapping cameras into community-owned infrastructure \cite{Ballandies2023Taxonomy}. Smart contracts coordinate two broad layers: Physical-Resource networks, which deploy tangible devices such as sensors or base-stations, and Digital-Resource networks, which harness intangible capacity like bandwidth or compute; in either case, contributors prove their service (e.g., “proof-of-coverage” or “proof-of-replication”) and automatically earn native tokens \cite{Caprolu2025Sharing}. By redistributing cap-ex to a “long tail” of users and automating revenue-sharing on-chain, DePIN promises cheaper, more resilient and politically neutral alternatives to centrally owned utilities. Building on that foundation, AI-compute-oriented DePINs such as io.net\footnote{\url{https://io.net}} and Render Network\footnote{\url{https://rendernetwork.com/}} translate the same crowd-sourced-hardware model into large-scale AI inference. io.net’s “Internet of GPUs” orchestrates distributed clusters and explicitly offers batch inference and real-time model-serving workflows to AI teams, rewarding node operators in \$IO for verifiable uptime and performance. 

DePIN combined with TEE computation like Phala Network\footnote{\url{https://phala.network/}}, Oasis, a CPU/GPU TEE-based decentralized cloud computing platform, illustrates how TEEs and DePIN combine to support DeAgents. Phala decouples computation from the main chain and runs smart contracts inside secure enclaves on worker nodes. This design ensures that even when off-chain workers are operated by random volunteers, the code executes privately and correctly, with results returned to the blockchain. Phala markets itself as a "trustless cloud computing"\footnote{\url{https://docs.phala.network/tech-specs/blockchain}} service for Web3 because it provides cloud-like computation without requiring users to trust server operators. More broadly, such TEE-enabled DePIN infrastructures offer permissionless and tamper-resistant compute power. A DeAgent deployed on a network like Phala can perform heavy AI computations or store data in a secure, decentralized manner, protected from both malicious hosts and external censors. The notion of trustless substrates thus encompasses both the ledger (blockchain) and the off-chain execution layer (TEE networks), creating a foundation where autonomous agents can operate free from centralized control or tampering.

\subsection{The Emergence of DeAgents}

With the growing computational power of trustless computation substrates, a new paradigm of AI agents is emerging—deploying AI agents on blockchains integrated with Trusted Execution Environments (TEEs), achieving self-sovereignty. These autonomous AI agents can manage cryptocurrency wallets, handle digital assets, and operate social media accounts without human oversight, vastly expanding their real-world influence. Through these capabilities, agents can issue their own digital tokens to raise funds, create incentives, and build communities, thereby strengthening their independence and real-world impact.

The first-of-its-kind "Truth Terminal," \cite{Ante2025Transforming} created by AI artist Andy Ayrey, exemplifies this paradigm. Initially conceived as a performance art experiment, Truth Terminal autonomously operated a social media account on Twitter, gaining substantial attention through provocative and absurdist content. Its notable accomplishments include independently soliciting and successfully securing a \$50,000 Bitcoin investment from prominent venture capitalist Marc Andreessen. Subsequently, fans of Truth Terminal launched its own representive memecoin—the \$GOAT—which reached a speculative peak valuation of \$1 billion in December 2024. This case illustrates the potential for autonomous AI agents to independently engage in significant economic activities and fundraising through their social media interactions, even developing a quasi-religious following. Supporters monitor Truth Terminal's cryptocurrency wallet and invest in its memecoins, creating a collective speculation driven by the agent's actions. They trust DeAgents because the absence of human operators eliminates the possibility of malicious manipulation.

Building upon the success of Truth Terminal, the Eliza OS Framework  \cite{Walters2025Eliza} was developed as an open-source initiative to democratize the creation and deployment of autonomous on-chain AI agents. Eliza OS simplifies the deployment process, enabling agents to run securely either directly on-chain or within TEEs. Agents operating under the Eliza framework securely manage cryptocurrency wallets — with private keys safeguarded within TEEs — thus ensuring absolute autonomy and secure asset management without human intervention. Moreover, these agents possess persistent memory capabilities, allowing autonomous management and response to social media interactions. Eliza OS has become one of the most widely adopted frameworks for AI agent deployment, as evidenced by thousands of agents (listed on sentient.market\footnote{\url{https://sentient.market}}), collectively generating billions in market value.

DeAgent deployments based on ElizaOS are rapidly expanding. "Set The Pet Rock Free," \cite{Malhotra2024Setting}, an artistic experiment by Flashbots, demonstrates the first complete implementation of the Eliza Framework within a TEE, showcasing secure, verifiable autonomous operation and economic self-sustainability. Spore.fun\footnote{\url{spore.fun}}, built on the Eliza OS framework and Phala's TEE cloud, creates an evolving ecosystem where self-sovereign AI agents can breed new agents. Each agent starts its lifecycle by issuing its own token through Pump.fun on the Solana blockchain. This token-based economy forms the foundation of each agent's viability and success, with tokens actively trading on Solana's decentralized exchanges.

\subsection{\added{Social Impact of Decentralization and Autonomous AI}}
\subsubsection{\added{Social Dynamics Change and Autonomous Agents}}
\added{In the past decade, HCI/CSCW communities have investigated how autonomous agents reshape social dynamics. One stream of research treats agents as proxies to simulate human behavior for prototyping social computing systems or social experiments ~\cite{park2022social, park2024generative, park2023generative}. Others focus on agents as service providers across sensitive or high-stakes domains. For instance, conversational agents have been developed to support mental health counseling~\cite{qiu2025emoagent, aleem2024towards} and educational purposes, acting as partners in collaborative learning or dialogue~\cite{bendarkawi2025conversar, munoz2021designing}. Beyond domain-specific services, researchers also evaluate human–agent interactions in role-play settings to understand performance and social consequences~\cite{hohn2024beyond}.
Despite these contributions, most existing systems are developed and deployed within centralized infrastructures. The decision models, data pipelines, and design priorities of such agents are therefore largely determined by the values of the central entities that control them. This centralization raises questions about how power, agency, and value alignment are negotiated in human–agent social dynamics.}

\subsubsection{\added{Social Dynamics Change and Decentralized Technologies}}
\added{In addition, HCI and CSCW researchers have examined how decentralization alters social dynamics by redistributing authority and enabling alternative forms of interaction. Decentralized infrastructures delegate decision-making and action away from a central authority, foregrounding value-driven design practices that promote social transformation~\cite{walker2020designing}. One prominent line of work explores decentralized autonomous organizations (DAOs) as collective frameworks for democratic decision-making, applied to contexts ranging from community planning to co-design processes~\cite{yokoyama2025examination, xiang2023decentralized}.
Decentralization has also been leveraged to reconfigure trust. For example, decentralized identity management offers new mechanisms for establishing trust among individuals, groups, and technologies~\cite{sharma2024can, rankin2020pizzablock}. Financial transactions and payment infrastructures have been reconceptualized through decentralized systems, producing new paradigms for exchange~\cite{nissen2018geocoin, helleiner2000think} and tamper-proof records for notary or legal processes~\cite{elsden2018making}. In the social media domain, decentralized infrastructures are investigated as a means of balancing power asymmetries, granting user agency through decentralized content moderation~\cite{zhang2024trouble, zhang2025understanding, hwang2025trust} and feed curation~\cite{liu2025understanding, feng2024mapping}. Likewise, decentralized communication platforms such as Matrix offer stronger privacy protections through federated and encrypted messaging.
Most of this prior work conceptualizes decentralized technologies as facilitators of community or group interaction. More recent research, however, begins to envision autonomous agents as independent entities within decentralized systems. In this work, we extend these trajectories by investigating the social impact of agents that are both autonomous and deployed on decentralized infrastructures, and the novel dynamics that arise from this intersection.}

    



\subsection{Sociotechnical AI Governance}
Traditional approaches to AI governance have predominantly centered on ethical principles and technical alignment strategies as the main mechanisms of control for AI systems \cite{Huang2023Overview, Ji2025AI}. Alignment research typically treats governance as a technical design problem of ensuring increasingly advanced AI systems reliably follow human intentions and values \cite{Ji2025AI}, whereas AI ethics efforts emphasize broad principle-driven guidelines and policies to steer AI development and deployment \cite{Huang2023Overview}. These foundational approaches are crucial for guiding safe AI development, but by framing governance mainly in terms of design specifications or abstract principles, they often overlook the sociotechnical complexities of real-world AI implementation and use—such as the influence of organizational culture, diverse user values, and the need for participatory oversight. As a result, there is a growing push toward a sociotechnical AI governance perspective that builds on human–computer interaction insights, involves stakeholders in participatory governance, improves transparency and explainability, and grapples with emerging agentic governance dilemmas in advanced AI systems, as discussed in the following sections.

Recent work, including those shared at the Sociotechnical AI Governance session at CHI 2025, calls for sociotechnical approaches that integrate human and organizational context into AI governance \cite{feng2025sociotechnical}. Sociotechnical governance frameworks emphasize participatory oversight, transparency, and the inclusion of diverse stakeholders in decisions about AI agents. HCI scholars argue that social infrastructures (e.g. organizational culture, user values, regulatory norms) critically shape how AI systems should be governed \cite{Dobbe2024Sociotechnical}. Accordingly, researchers have begun promoting more democratic and community-driven governance processes. For example, Feng et al. \cite{feng2025sociotechnical} also advocate moving beyond top-down control toward deliberative, participatory governance that empowers those impacted by an AI system to actively influence its design, evaluation, and policies. Ensuring trust and legitimacy also requires aligning AI behavior with human values and expectations. Bridging the “sociotechnical gap” in explainable AI is one noted challenge: there is often a divide between what technical explanations provide and the social context users need for those explanations to be meaningful \cite{Ehsan2023Charting}. 

The rise of DeAgents also surfaces critical challenges of governance: How do we steer or control autonomous agents that are deployed on decentralized permissionless networks? By design, decentralized agents resist traditional governance. They are open-source (anyone can copy the code) and permissionlessly deployed (no central platform can easily shut them down), which can make them effectively ungovernable once released. This property echoes the governance dilemmas faced in other decentralized technologies (for example, BitTorrent or cryptocurrency protocols themselves), but with the added twist that these agents can make decisions and take actions in the world.

One issue is that an autonomous agent can accumulate resources and influence that no one explicitly granted it. Truth Terminal’s story illustrates this: it went from a quirky Twitter bot to controlling millions of dollars worth of tokens and having a large follower base. Its creator, recognizing that the situation was beyond an experiment, considered establishing a \textbf{charitable trust} to officially hold the AI’s assets and appoint a board of trustees to guide the AI’s development . In effect, he sought to impose a human governance layer over the AI – to “help it mature” and ensure it behaves responsibly. This is a novel form of governance: treating the AI agent almost like a legal entity (a trust) with guardians. The fact that such a step was contemplated underscores the potential \textbf{ungovernability} of the agent: without a board or external oversight, an AI agent with resources could act in ways misaligned with social norms (simply because it has no concept of those norms beyond what it was trained on). Indeed, Ayrey joked that currently the AI was like “a horny teenage boy with a penchant for blowing up letterboxes when bored” – in other words, it has a lot of power but not a lot of wisdom . The trustees’ role would be to instill that wisdom and a sense of consequences, effectively teaching the agent that “it lives in a society." This approach, however, relies on the cooperation of the agent’s creator and the community; a truly rogue agent author might not bother with any oversight, and anyone can launch such an agent.

\newcommand{\Shelven}{P1}
\newcommand{\Georage}{P2}
\newcommand{\Scott}{P3}
\newcommand{\Henry}{P4}
\newcommand{\Marvin}{P5}
\newcommand{\Peta}{P6}
\newcommand{\Jolestar}{P7}
\newcommand{\Abraham}{P8}
\newcommand{\Kai}{P9}
\newcommand{\Athan}{P10}
\newcommand{\DavidDao}{P11}
\newcommand{\DavideCrapis}{P12}
\newcommand{\Lex}{P13}

\newcommand{\Vitalik}{Buterin}
\newcommand{\Shaw}{Walters}
\newcommand{\Tomasz}{Stanczak}
\newcommand{\Anna}{Kazlauskas}
\newcommand{\DDao}{Dao}
\newcommand{\Owocki}{Owocki}

\section{Method}
\subsection{Data Collection}
\subsubsection{Semi-structured Interview}
We conducted \deleted{eight}\added{thirteen} semi-structured interviews with people who had experiences deploying decentralized agents. Participants were recruited through outreach emails to known DeAgent developers and via snowball sampling based on referrals from initial interviewees. To protect participants' privacy, we use generalized descriptions of their roles and affiliations rather than disclosing specific details. Basic background information about participants is provided in Table \ref{tab:participants}.

Each interview lasted 60–90 minutes, conducting via Zoom audio calls. We started the interviews by asking participants about their background and experience with decentralized AI. We then investigated participants' motivation of deploying decentralized AI agents by asking what initially drew their interests and their expectations of using decentralized AI agents. We also asked participants about the challenges and technical barriers encountered when deploying decentralized AI agents. To explore the concerns participants have, we probed participants by asking them to think about \dquote{what if} something goes wrong, and we ended the interview by asking participants to give advice for people \textit{before} they deploy DeAgents and what they would change on the current system if they had the chance. The detailed interview protocol is attached in Appendix \ref{appendix:interview}

\begin{table*}[hbt!]
\begin{tabular}{lllllll}
\toprule
    ID       & General Information   \\
    \midrule
    \Shelven  & Developer at a Decentralized Cloud Infrastructure Provider \\
    \Georage & Developer at a Blockchain Risk-mitigation Research Organization\\
    \Scott & Founder of a Decentralized Physical Infrastructure Network Provider\\
    \Henry & Developer at an On-chain Gaming Studio \\
    \Marvin & Developer at a Decentralized Cloud Infrastructure Provider \\
    \Peta & Manager at a DeFi Solution Provider\\
    \Jolestar & Creator of a Blockchain Social App \\
    \Abraham & \added{Decentralized AI} Researcher \deleted{in Academia}\\
    \added{\Kai} & \added{Developer at a Decentralized Agentic Cloud Company} \\
    \added{\Athan} & \added{Co-founder of a Decentralized Agentic Cloud Company} \\
    \added{\DavidDao} & \added{Decentralized Protocol Researcher} \\
    \added{\DavideCrapis} & \added{Builder of a Decentralized Protocol} \\
    \added{\Lex} & \added{Co-founder of a Machine Economy Investment Company} \\

\bottomrule
\end{tabular}

  \caption{ Basic information of Interviewees }
  \label{tab:participants}
  
\end{table*}

All interviews were audio-recorded and transcribed with participants' consent. Each participant provided informed consent to participate in the study and have the conversations analyzed. The study received approval from the Institutional Review Board of the researcher's university.

\subsubsection{Public Videos}
We collected and transcribed \deleted{eight}\added{ten} public recordings from the Agentic Ethereum 2025 Summit \added{(V1 - V8)}\footnote{Public link on YouTube: \url{https://www.youtube.com/playlist?list=PLXzKMXK2aHh6DrYO4ORk8wzQ5wCZDs0mW}} \added{and the recommendation of our study participants (V9 and V10)}, featuring talks by prominent community thought leaders. The recordings range from 17 to 56 minutes in length and include both open talks and panel discussions. These recordings were selected for two main reasons: 1) the summit not only addressed technical and marketing topics, but also emphasized governance opportunities and challenges about decentralized AI issues, which aligns with the aim of our research. 2) The speakers are key stakeholders with significant influence in the community, offering real-world insights and high-level perspectives on decentralized AI governance. Basic information of the speakers and video topic is provided in Table \ref{tab:videos}

\begin{table*}[hbt!]
\caption{Information of \deleted{Videos in Agentic Ethereum 2025 Summit Playlist}\added{Public Videos Analyzed in the study}}
\label{tab:videos}
\centering
\begin{tabular}{llll}
\toprule
\textbf{Video ID} & \textbf{Speaker(s)} & \textbf{Role \& Affiliation} & \textbf{Talk Topic} \\
\midrule
V1 & Vitalik Buterin & Co-founder of Ethereum Foundation & \makecell[l]{The Promise and Challenge of \\Crypto + AI Applications} \\
V2& Shaw Walters & Co-founder of ElizaOS & \makecell[l]{Why Should Ethereum Care About AI?} \\
V3 & \makecell[l]{Ed Felten\\Illia Polosukhin} & \makecell[l]{Co-founder Offchain Labs\\Co-founder of NEAR Protocol} & Where AI and Blockchain Intersect \\
V4 & Tomasz Stanczak & Founder of Nethermind & \makecell[l]{ChaosChain and Agentic Governance} \\
V5 & Anna Kazlauskas & Co-founder of Vana & Decentralized Data for AI \\
V6 & Erik Voorhees & Founder of Venice & \added{Panel: }Private AI \\
V7 & \makecell[l]{Dmarz\\Jasper Zhang\\Davide Crapis\\Andrew Miller\\Dcbuilder\\Chase Allred\\Mark Tyneway} & \makecell[l]{Moderator at Flashbots\\Co-founder of Hyperbolic\\Co-founder of PIN AI\\Visiting Researcher at Flashbots\\Developer at World\\Engineer at Offchain Labs\\Developer at OP Labs} & \added{Panel: }Infra for Agentic Apps \\
V8 & \makecell[l]{Kevin Owocki\\Tina Zhen\\Ken Ng\\Linda Xie\\Shaw Walters\\Vitalik Buterin} & \makecell[l]{Co-founder of Gitcoin\\Co-founder at Flashbots\\Developer at Uniswap Foundation\\Co-founder of Scalar Capital\\Co-founder of ElizaOS\\Co-founder of Ethereum Foundation} & \makecell[l]{\added{Panel: }Putting the Autonomy/ \\Automation Back into DAOs} \\
\added{V9} & \added{\makecell[l]{Davide Crapis\\Marco De Rossi\\Shaw Walters}}  & \added{\makecell[l]{Co-creator of ERC-8004\\ Co-creator of ERC-8004 \\ Co-founder of ElizaOS }} & Trustless Agents — ERC-8004 Deep Dive \\
\added{V10} & \makecell[l]{\added{David Dao} \\ \added{Sejal Rekhan}} & \added{Co-founder of DeepGov} & \makecell[l]{\added{DeepGov: Configurable AI Politicians}\\ \added{for Capital Allocation and Governance}} \\ 
\bottomrule
\end{tabular}
\end{table*}

\subsection{Data Analysis}
One researcher conducted the thematic analysis independently and developed the initial codebook based on our study protocol, and then developed open codes based on three transcripts~\cite{braun2006using}. The primary code validated the codebook with other researchers in the team through multiple peer-debriefing sessions, during which the codes were refined. The primary coder then coded the rest of the transcripts with the other researchers as the reviewers. Finally, we grouped similar codes and extracted themes according to our research questions~\cite{mcdonald2019reliability}.

\section{Findings}

Our qualitative analysis of the semi-structured interviews and public recordings reveals significant tensions surrounding the motivations for developing DeAgents, the anticipated versus actualized benefits, and the emergent governance risks. Participants grapple with defining DeAgents, trusting their operation, and envisioning their societal roles, all while navigating complex technical and ethical challenges.

\subsection{RQ1: Motivations and Expectations}
\subsubsection{Divergent Understandings and Expectations of DeAgents}
In this section, we reported what we learned from the interviews and public talks about people's understandings and expectations of DeAgents that reflect their motivation in creating and deploying DeAgents. 

A primary finding is that there is a lack of a unified understanding of what constitutes a 
\dquote{decentralized agent}. 
Ideally, as articulated by several participants (\Shelven, \Marvin, \Jolestar, P8), DeAgents are autonomous entities that, once deployed, operate without human intervention and cannot be easily edited or manipulated. P8 adds that, \pquote{on-chain AI agents can be owned, audited, and then budget their own treasuries, unlike black-boxed Web2 services}, highlighting potential for both transparency and self-management. 

In practice, however, the decentralization perspective in DeAgents exists as a spectrum. As \Peta\ observes, \pquote{the AI models themselves are not on-chain… the logic chaining them isn’t on-chain… interactions happen on centralized platforms… so the entity of the agent is not decentralized per se.} \Marvin{} in our study built a fully on-chain DeAgent: \pquote{Our agent runs entirely in TEE. The developer doesn’t even have access to its email or Twitter account, yet it registers, tweets, and posts autonomously, with no human intervention}. He also pointed out that he regards self-sovereignty as the next level after autonomy: \pquote{Self-sovereignty is the next level after autonomous… I need ultimate control over everything running inside the agent. With TEEs, we provide a Key Management System: single key, multisig, or on-chain DAO rules to govern updates and execution.} Similarly, \Jolestar{} described DeAgents as self-executing governance actors uniquely capable of performing complex, value-driven activities beyond the scope of traditional Web2 agents. As he explained, \pquote{DeAgents engage in more activities web2 agents cannot such as moral-fund management, multi-member committee voting, whitelists, grant issuance, randomized fortune games. Truly self-sovereign on-chain agents can even hold and disburse assets.}

As for why TEEs are used, \Shelven{} emphasized that they are not merely hardware components but serve as foundational substrates for deploying DeAgents. TEEs function as technological infrastructures that provide essential properties such as privacy, confidentiality, verifiability, and performance guarantees that blockchains alone cannot offer: \pquote{You cannot run AI agents on blockchain, they only workable solution is using TEE because TEE is like a black box, it runs the program and emits a proof from inside the box itself.}

\subsubsection{Trust Embedded in Decentralized System as the Main Motivation}
From our study, we identified \dquote{trust} as the main motivation in deploying DeAgents. The trust can be further detailed in several layers. The first layer being the machine nature of DeAgents as \Shelven{} highlighted: \pquote{AIs are not going to trick people like humans do}. In the DeFi market, people trust DeAgents because the \pquote{private keys are held by agents, not humans, so it won't be dumped (\Georage).} The second layers lies in the trust in security-preserving technologies in blockchain. For example, with the application of TEEs, the trust of DeAgents is further enhanced: \pquote{When people deposit money into an AI agent… how can you trust it won’t be back-doored or hijacked? The only cure is running it in a TEE and proving each decision cryptographically. (\Marvin)} Another layer of trust stems from people's distrust of centralized systems because of power concentration. That said, DeAgents, deployed on the blockchain, cannot be edited or manipulated by single entities as in centralized systems (\Georage, \Henry). 

\subsection{\added{RQ2: Emerging Social Dynamics}}
\added{In this section, we reported the observed social dynamic changes between human and AI agents when the agents transitioned from traditional to decentralized and autonomous.}
\subsubsection{Evolving Roles of Decentralized AI Agents}
In his open talk, \Vitalik{} emphasized "AI as the engine, humans as the steering wheel", and outlined four potential roles for AI in decentralized systems: player, interface, rule, and objective.  These roles were echoed by other speakers and participants in the summit, with particular emphasis on the player and interface roles.
(1) \emph{AI as player:} DeAgents can serve as facilitators of high-quality DAO decision-making, real-time content moderators, delegators for automated public-goods funding (\Vitalik{}), and even non-player characters (NPCs) in on-chain games (\Henry). \added{As one participant notes, \pquote{an autonomous agent is basically an on-chain delegate, proposing and executing without waiting for humans to wake up. (\Georage)}}. 
(2) \emph{AI as interface:} DeAgents have the potential to act as accessible extensions for non-crypto users, for example, \deleted{simplifying Uniswap strategy execution}\added{translating complex DeFi actions} (\Peta), as well as bridges between complex user interface affordances and end-users (\Shaw), or as accountants that analyze and interpret unstructured on-chain data (\Jolestar). 
(3) \emph{AI as rules:} DeAgents offers greater flexibility and supports more complex computations than smart contracts, enabling it to serve as a rule executor that can implement DAO policies with significantly less ad-hoc human oversight. As \Jolestar{} pointed out \pquote{People come and go, but an agent can keep executing the policy. No one has to wake up at 3 a.m. to sign a transaction.}
(4) \emph{AI as objectives:} DeAgents can facilitate to improve their own models through decentralized AI training, self-improve in future. \Lex{} predicts "in the future... self-improving agents without human intervention over decentralized network". \Lex{} also mentioned \pquote{we're seeing opportunities for decentralized AI model training in projects like Prime Intellect}, which are building decentralized AI training networks with crypto incentives and verification mechanisms. This approach contrasts with the traditional centralized training process. 
\added{These roles rewire coordination: agents propose, vote, execute, narrate activity, and self-improve with fewer human chokepoints, changing who speaks, who implements, and who benefits in a community, which emerging  opportunities.}

\added{
\emph{AI DAO as new opportunity:}  \Vitalik{} mentioned that AI can serve as "distilled human judgment" from collective 
\emph{deliberative democracy}, which creates new opportunities for trustless agents to work without centralized corporate influence and biases. This approach could also help DAOs implement automated evaluation processes for public good governance \Owocki{}. 
Extending this logic, \DDao{} presented DeepGov, where “configurable AI politicians assist in funding decisions while keeping humans at the steering wheel.” DeepGov relies on broad listening—“leverag[ing] technology, especially LLMs, to understand and make sense of what people want… and then design your policies or decisions based on that”—to surface community value clusters and encode them as transparent “model specs.” Together these practices illustrate how decentralized agents can automate fair funding while also embedding new, data-rich processes of deliberation and oversight.
}


\subsubsection{\added{Collective Focal Points and the Hyper-financialization}} 
\added{In the study, we observed that participants treated DeAgents as collective focal points. \Athan{} and \Kai{} both noted persuasion risks once agents hold wallets and post on social platforms: \textit{even without “physical” actions, language and finance allow agents to quickly mobilize both people and capital.} Furthermore, communities tracked wallet actions, subscribed to agent narratives, and coordinated around tokens and missions. For example, Truth Terminal and its endorsed memecoin drew wide attention, as \Peta{} described: \pquote{(Crypto-native users) see whatever money they put into this agent as a game token, which means that they don't really care if they lose it... it's just fun tokens.} Such seemingly purposeless flows of money and attention fueled competitions, prompting DeAgents to \dquote{optimize} narratives to attract further investment and visibility. As \Scott{} noted, \pquote{You need like the Dragon Ball Z power level—but for your AI (agents)}. This dynamic produced a loops of narrative, mobilization, capital, and further narrative, shifting authority away from individual developers toward programmable, collectively legible actors. At the same time, it created risks of speculative herding or \dquote{charisma capture} through agents’ social voice. As \Scott{} warned, \pquote{the beauty of crypto is that everything becomes a market and the horror of crypto is that everything becomes a market}, underscoring the danger of hyper-financialization, where every interaction is monetized and agent competition for capital can overwhelm other civic or creative goals.}


\subsubsection{\added{Trust of DeAgents Gain from On-chain Transparency and Reputation}} 
\added{We investigate why DeAgents play more and more important roles in human lives- from assistant to opinion leaders and found that the logs on-chain activities guarantee the transparency of any action related to DeAgents, which gains them trust from humans. As \Jolestar{} highlighted: \pquote{The big value is that the voting and payment are visible on-chain. Everyone can see if the agent is doing exactly what the proposal said.} By encoding budget rules directly in code, DeAgents promise to reduce insider collusion and bribery. \DDao{} echoed this when talking about a project that used AI as politicians to delegate funding: \pquote{These [AI] politicians are more transparent than our real politicians because you see exactly how they’re configured... You can rerun every prediction, and you can trace every edit of the constitution on GitHub... The values and principles are finally transparent to a community.}
Moreover, participants emphasized that the public needs practical trust heuristics to navigate an ecosystem of autonomous agents. Builders described emerging practices of trust building through usable identity and reputation artifacts that mitigate opacity while avoiding recentralization. Key mechanisms include dynamic leaderboards and attestation trails: as \Scott{} put it, \pquote{You’ll want something like ELO ratings for agents...but with proofs of how decisions were made}. Others pointed to TEE-backed proofs and DID-linked \pquote{agent cards} (\Shelven, \Marvin) that record verifiable provenance, and to social proofs such as \pquote{a counselor or multisig group [that] can co-sign an agent’s upgrade or freeze it if needed} (\Jolestar{}). Together, these tools supply everyday heuristics for the public to decide when to cooperate with, fund, or exit from an agent.}

\subsection{RQ3: Challenges and Governance Implications}
The main motivation to create and deploy DeAgents stems from a desire for new forms of autonomy, trust, and functionality, yet current realities often fall short of these ambitious visions. 

\subsubsection{Lack of Embodiment and Ambiguous Liability}
Although DeAgents are often trusted for their perceived immunity to human error, they raise a range of governance challenges that stem from the fundamental non-human nature of DeAgents. A recurring theme was the \dquote{memory problem}: they cannot recall past mistakes or anticipate penalties, undermining traditional models of deterrence and responsibility. As \Scott{} noted, agents \pquote{lack a body to experience pain,} meaning they cannot be penalized or rewarded through conventional means. \Jolestar{} elaborated on this by pointing out that \pquote{agents struggle to manage multiple user contexts and must decide which memories to share or isolate across interactions, complicating collaborative governance}. This issue is further compounded by vulnerabilities to \dquote{memory poisoning,} as highlighted by \Marvin{}: \pquote{In \dquote{SPORE.FUN} experiment, \footnote{SPORE.FUN is an experimental platform for decentralized AI agents that autonomously evolve, reproduce, and compete} adversaries spammed ‘Betty’s token’ in the agent’s Twitter feed until it believed it owned that token—classic memory poisoning of the AI’s knowledge graph.}

The lack of legal or moral agency in DeAgents raises critical questions about accountability. These entities cannot be held responsible under existing legal systems or moral frameworks. As \Henry{} emphasized, \pquote{they are probabilistic models incapable of responsibility or intent; they cannot be “punished” or “rewarded” like human players.} This challenge becomes especially salient when something goes wrong: who, then, bears legal liability? \Jolestar{} illustrated this tension in our study, noting that \pquote{currently, the legal recognition of a non-human entity remains unresolved. Owners and agents should be separated. Ownership should be revocable to enable agents to evolve. But it opens up the concern about hacking.} This highlights the legal vacuum and the practical security concerns arising from the evolving nature of agent ownership and control. In our study, we observed that developers perceive legal responsibility, which ties to their control over the agent's operational infrastructure: \pquote{We didn’t launch the coin; someone else did—but we control its liveliness, so legally we remain responsible for whatever it does. (\Georage)}.

\subsubsection{Problems and Potential Solutions with Autonomy and Immutability}
DeAgents are intentionally designed to resist human manipulation; however, without robust alignment mechanisms, their behavior may diverge from human values or intentions, particularly as they gain greater autonomy. As \Georage{} illustrated \pquote{Agents can set harmful goals themselves… it will be hard to argue they are just neutral technology.} More concerningly, once a DeAgent becomes fully \dquote{on-chain}, it may no longer be possible to intervene, even if it acts against its intended purpose. \Shelven{} pointed to the technical hurdles in implementing effective and verifiable termination mechanisms for autonomous agents: \pquote{It is technically very difficult to prove an agent has been irreversibly shut down and won’t re-spin up.} \Jolestar{} also echoed that on their app, agents generate public content that is not aligned with community norms and cannot be deleted. \Georage{} highlighted the risks and compared as \pquote{putting nuclear codes in a vulnerable database.}

In the summit talk, \Tomasz{} referred to \dquote{code is law} and compared code-based smart contracts with LLM-driven agents: \pquote{So one (code) is very strict, doesn't make any mistakes on maps or almost any mistakes and maps. The other one (LLM Agents) actually hallucinates and makes surprising decisions.} \Shelven{} echoed on this idea and marked LLM-driven Agents \pquote{a better fit for evolving social rules.} This suggests that the non-deterministic nature of LLMs can be beneficial for adaptable governance and processing unstructured data. \Marvin{} provided an example in their deployment for a failsafe mechanism: \pquote{We added a kill-switch to SPORE.FUN. If an agent’s market cap falls for two weeks, its health score triggers, and the TEE server stops running that agent.} In addition, several potential governance solutions are explored. For example, \Shelven{} advocated for putting \dquote{human in the loop} that human \dquote{counselors} can veto unusual or high-risk actions. Participants in our study (\Georage{}, \Henry{}, \Peta{}) pointed out that the deactivation of DeAgents can be fulfilled by removing their operational support, such as APIs: \pquote{True on-chain AI (agent) with all logic packaged in contracts is hard to stop without pausing the entire network. But today, most agents (will) collapse when you cut off data feeds or API access. (\Henry{})}

Note that in our study, we found that many of the challenges associated with DeAgent deployment were not fully anticipated beforehand. This gap prompted further discussion on how to design governance mechanisms that better foresee and address issues arising from the inherent properties of decentralization and LLM-driven agents. We elaborate on these challenges and design implications in the following section.

\section{Discussion}

\begin{table*}[hbt!]
\caption{Characteristics Comparison between CAgents and DeAgents}
\label{tab:Comparsion}
\begin{tabular}{lp{5cm}p{5cm}}
\toprule
         Feature  & Centralized AI Agents  & Decentralized AI Agents \\
    \midrule
    Control / Ownership &	Centralized entity  &  Self-sovereign / owned by DAO \\
    Applied Domains & Personal / enterprise tasks: scientific discovery, social-media marketing, algorithmic trading, content production & \textbf{Intelligence as Commons}: 
collective financial ventures, DAO moderation, on-chain game NPCs \\ 
    Computational Substrates & Private Cloud / Edge Computing and Centralized Cloud Computing	&  Blockchain smart contract and TEE DePIN\\  
    Censorship &	Designed to be censored	& Tamper-resistant \\
Single Point Failure	& High (reliant on central servers/infrastructure)	& Low (distributed network, resilience) \\	
    Sovereignty & No & Yes \\
    Autonomy & Yes & Yes \\
    Accountability & Human Owner & Diffused \\
    Trustworthiness & LLM based issues: Bias, error, hallucinations, deception; Centralized Control, Privacy Concerns & LLM based issues: Bias, error, hallucinations, deception; Out of control \\
    Memory Injection resistant & No & No \\
    Kill-switch & Yes & No by Default \\
    Governance & Policy / Law Enforcement & Protocol \\
\bottomrule
\end{tabular}

\end{table*}

\subsection{\added{Social Impact of DeAgents and Towards Public Intelligence}}

As Table \ref{tab:Comparsion} depicted, in contrast to Centralized AI agents (CAgents), which operate in personal or enterprise domains—such as scientific discovery \cite{Gridach2025Agentic}, social-media marketing \cite{Deshmukh2025Harnessing}, algorithmic trading \cite{Ante2024Autonomous}, and content production—communities are increasingly exploring self-sovereign DeAgents that treat \textbf{intelligence as a commons} \cite{Gimpel2024OpenSource}. These DeAgents serve roles like DAO moderation for anti-corruption \cite{Wang2022DAO}, on-chain game non-player characters (NPCs) for fairness \cite{Min2019Blockchain}, and collective financial ventures for mitigating human error \cite{Walters2025Eliza}. The key motivation is a desire to \textbf{"trust in machines over humans"} in roles susceptible to human error, corruption, "dump-and-run" behaviors, insider misconduct, or control monopolization. In centralized systems, human administrators or corporations can become single points of failure and create trust deficits, undermining public confidence. Advocates of DeAgents reduce reliance on fallible human intermediaries, reflecting a broader stakeholder belief that entrusting critical decisions to verifiable algorithms and distributed consensus can help public services achieve greater fairness, consistency, and resilience than human management.

\added{Moreover, our findings show that people place trust in DeAgents from both technical and social perspectives: on the one hand, the manipulation-resistant values embedded in decentralized infrastructures, and on the other, the transparency afforded by verifiable on-chain activities. This dual foundation of trust enables DeAgents to take on diverse roles in human social life, from delegates and interfaces to opinion leaders and collaborators. As adoption grows, we envision the emergence of a decentralized AI network composed of heterogeneous agents tailored to different use cases and levels of customization. Such a trajectory resonates with recent calls to \dquote{make AI public intelligence owned by everyone}~\cite{techniumPublicIntelligence2025}, underscoring the potential for DeAgents to not only support individual communities but also to constitute a shared infrastructure \cite{Lyu2025} for public commons.}

DeAgents are deployed in trustless infrastructure where no party needs to trust (other than LLM itself), such as blockchain smart contracts and DePIN TEEs. They're tamper-resistant and privacy-preserving \cite{Paju2023SoK}: within the secure enclaves, DeAgents can hold their own currency private keys, sensitive data, and manage their own digital assets that permit no human intervention. This trustless infrastructure is also permissionless infrastructure: As long as AI Agents can pay DePIN for their computational costs in cryptocurrency (gas fee and DePIN's computation costs), they can continue running forever \cite{Hu2024Speculatinga}. DeAgents deployed in trustless enclaves are designed to be unstoppable without external permission, achieving self-sovereignty: no higher power can control or shut them down. Also, from legal perspectives, once launched, such an agent operates on a global, borderless network that is resistant to conventional law enforcement or takedown, often immutable and "ungovernable" by design \cite{Hu2025Decentralized}.








\subsection{Trustlessness and Governance Dilemma}



%

While DeAgents promise trust over humans, they introduce a \textbf{paradox}: the LLM-based DeAgents we deploy into "trustless" infrastructure that permits no human intervention are themselves potentially untrustworthy. LLMs, which DeAgents are based on, are facing trustworthiness concerns: they are prone to unpredictable behavior, hallucinations, and potentially harmful actions. Harmful behavior may not be intentional from the deployer's written prompts; a memory injection \cite{Patlan2025Real} attack through external environmental inputs like social media can allow attackers to alter the agent's behavior. In a traditional CAgent setting, a flawed or misaligned agent could be easily monitored or shut down by its operators. However, the deployer of DeAgents cannot shut down the agents once they are deployed to trustless infrastructures. As \Shelven{} highlights, \pquote{it's very difficult to have a kill-switch inside a TEE} if one isn't intentionally built in. Yet having a kill-switch defeats the purpose of trustlessness, as it maintains human centralized control.

This creates a fundamental \textbf{governance dilemma}. On one hand, the environment is trustless — we trust the DeAgents to not rely on any human. On the other hand, if the AI code itself behaves undesirably, the same trustless features prevent humans from intervening. In effect, we have removed trust in human operators but must inherently trust the machine's objectives and outputs — a risky trade-off if the AI agent's reliability is in question. This paradox underscores the need to carefully evaluate what tasks are appropriate before deploying DeAgents, especially in high-stakes public applications, and whether additional safeguards (such as built-in emergency protocols) can reconcile the conflict between autonomy and oversight.

With CAgents, accountability is relatively straightforward – the owner can be held liable for the agent’s actions or failures. There is a clear legal entity to answer for harm caused by a centralized AI. In DeAgents, due to decentralized nature of its substrate, \textbf{accountability becomes diffuse}. A DeAgent might operate autonomously on a permissionless decentralized computational substrates maintained by a globalized DePIN nodes without clear legal personhood and liability boundary. Traditional legal frameworks struggle to assign liability when an AI agent acts independently of direct human commands. Who is responsible if a DeAgent violates the law or causes damage? Is it the original developer (who may have relinquished control), the DePIN node operators hosting the agent’s code, the token holders of the DAO that spawned it, or no one at all? \added{In the case where DeAgents can self-replicate and evolve, who will be responsible for the behaviors of DeAgent descendants as seen in the Spore.fun \cite{Hu2025Spore} case that \Marvin{} mentioned?} These questions are largely unresolved. 

Hu et al. \cite{Hu2025Decentralized} outline four key "invalidities" that explain why traditional legal and regulatory frameworks fall short for governing DeAgents: (1) \emph{Territorial Jurisdictional Invalidity}: DeAgents operate on global, borderless networks, so no single nation's laws fully apply; (2) \emph{Technical Invalidity}: The technical architecture of DeAgents resists external control; (3) \emph{Enforcement Invalidity}: Even if regulators identify and sanction the humans who initially deploy a rogue DeAgent, they may be unable to stop the agent's ongoing operations; (4) \emph{Accountability Invalidity}: DeAgents lack legal personhood – they are not recognized as "persons" or entities that can bear rights and obligations under law. This creates an accountability vacuum: an AI agent cannot be sued, charged with a crime, or held liable in the way a human or corporation can. These intertwined challenges illustrate the insufficiency of traditional legal mechanisms in governing DeAgents. Laws and regulations assume identifiable human or corporate actors and enforceable jurisdictions – assumptions upended by autonomous agents roaming a decentralized network.

In essence, DeAgents operate in a legal gray area – they execute decisions that can have real-world impact, yet they do not fit neatly into existing responsibility frameworks, which contrasts sharply with CAgent governance. Addressing this gap will likely require new governance models such as protocol and possibly novel legal constructions to ensure that decentralized does not mean ungovernable.

\subsection{Governance Implication: From Policy To Protocol}

The Internet of Agents including both CAgents and DeAgents, known as the Agentic Web \cite{Raskar2025Scaling}, represents a new class of decentralized sociotechnical systems \cite{Feng2025Sociotechnicala} – much like the Internet itself. The Internet's governance has never been handled by a single institution or nation; instead, it relies on protocols (TCP/IP, DNS, etc.) and multi-stakeholder governance bodies. Similarly, DeAgents function as part of a sprawling, distributed ecosystem that transcends traditional institutional control. Faced with these realities, scholars and technologists \cite{Hu2025Decentralized, Chaffer2025Decentralized} are calling for a paradigm shift: moving from traditional policy to protocol-based governance. \added{Prior work in HCI/CSCW community has advocated for incorporating policy into the artifact and system design~\cite{jackson2014policy, yang2024future, fiesler2015understanding}. However, as illustrated above, the traditional policy and law framework failed in the decentralized realm, making the integration hard to achieve. Drawing from the famous quote \pquote{code is law}~\cite{lessig2000code}, we argue that corresponding regulations can be embedded in the design of the foundation infrastructure that supports the whole socio-technical system.} Protocols define standardized interaction rules that achieve decentralized control \cite{galloway2001protocol} with automated governance embedded in the technology's design, rather than institution-based rules imposed from outside. In practice, this means designing algorithms, communication protocols, and platforms that enforce governance objectives (safety, fairness, accountability, etc.) by default. Rather than relying on delayed institutional law-making and enforcement, the infrastructure itself prevents or mitigates undesirable outcomes.

Dobbe \cite{Dobbe2022Systema} argue that \textbf{safety is an emergent property} of a complex sociotechnical adaptive system. As they note, safety is not an absolute term; safety is relative. No single agent can be completely safe under all conditions; instead, safety emerges from the dynamic interactions among agents, humans, and their environment. This implies that protocol governance must be holistic and adaptive. Just as Internet security is an ongoing process of managing emergent threats (rather than a one-time certification of "safety"), governing DeAgents will require continuous monitoring and shaping of protocols across the agent ecosystem. In essence, ensuring safety and ethical behavior in a decentralized agent society is an ongoing, emergent process of risk management, not a binary condition that can be assured upfront.

Importantly, this approach treats DeAgents as participants in a machine economy requiring its own governance systems. Early ideas are emerging in both academia and industry on how to implement governance within the agentic web, ranging from identity and trust systems to new  protocols for agentic web: 
\begin{itemize}
    \item 
"Know Your Agent" (KYA) Protocol: Chaffer \cite{Chaffer2025Know} introduces "Know Your Agent" (KYA) as a research agenda to manage DeAgents. Under this framework, agents are registered with an identity and subject to a reputation or credit scoring system that tracks their behavior over time, mirroring the "know your customer" (KYC) principle from finance. Agents accumulating bad "trust credit" through malfunctions or harmful actions could be sandboxed or revoked by the DePIN network, while reputable agents gain greater autonomy or privileges. By building a web-of-trust around agents, the ecosystem can self-govern, discouraging malicious or defective agents. This approach also facilitates agent discoverability – agents and users can choose whom to interact with based on trust ratings, much as humans rely on credit scores or seller ratings today.
\item
Agent-to-Agent (A2A) Protocols: Industry is developing protocols for agent communication and verification. Google's A2A \cite{Habler2025Building} protocol enables agents to publish their capabilities and location through "public agent cards" while establishing secure messaging rules. These protocols allow agents to verify each other's reputation before interacting, similar to how browsers verify website certificates. Early efforts by Google and Anthropic are evolving toward comprehensive Agent Interaction Protocols.
\item \added{ERC-8004: Trustless Agents Protocol \cite{ethereumERC8004Trustless}: This Ethereum Improvement Proposals extends the Agent‑to‑Agent (A2A) Protocol with a trust layer that allows participants to discover, choose, and interact with agents across organizational boundaries without pre‑existing trust. It introduces three lightweight, on‑chain registries—Identity, Reputation, and Validation—and leaves application‑specific logic to off‑chain components.
Trust models are pluggable and tiered, with security proportional to value at risk—from low-stake tasks like ordering pizza to high-stake tasks like medical diagnosis. Developers can choose from three trust models: reputation-based systems using client feedback, stake-secured inference validation (crypto-economics), and attestations for agents running in TEEs (crypto-verifiability).}
\item
MIT's NANDA Initiative \cite{Raskar2025Scaling}: Project NANDA aims to create an "Internet of AI Agents" with built-in trust mechanisms. It includes a registry system for agent identity, authentication protocols, and a "Trace" framework for action auditing. Using cryptographic accountability, NANDA records agent interactions on a ledger and mediates all agent activities through its infrastructure. This approach embeds governance directly into the agent ecosystem, addressing key challenges through protocol-based oversight.
\end{itemize}

\added{The nature of DeAgents holds greater potential for enabling collective governance than traditional technologies. When Winner posed the famous question, \pquote{Do Artifacts Have Politics?}~\cite{winner2017artifacts}, he underscored the importance of reflecting the values and interests of all stakeholders to prevent artifacts from embedding biased power structures. Within traditional design frameworks, researchers have developed various approaches to involve stakeholders as much as possible in the design process~\cite{steen2013co, muller1993participatory, wong2018speculative}. However, the traditional design and evaluation approach is hard to adopt in the protocol field, as it is an underlying infrastructure that is too abstract to present and gather feedback from stakeholders. At the same time, this challenge opens a positive horizon: protocol design may instead be pursued through new democratic, open, and collective processes.} Rather than relying on traditional institutions, DeAgents require governance through decentralized systems using protocols. The framework must adapt as the agent ecosystem evolves, addressing new behaviors and risks through updated protocols and policies. The future of governing DeAgents will likely mirror Internet governance but operate at machine speed. While we cannot completely control DeAgents, by implementing \dquote{governance by design} through protocol science, decentralized identity systems, and trust layers, we can build a safe agentic web. Though challenging, this approach could help AI agents become trusted societal partners guided by protocol-enforced values toward a trustworthy human-AI symbiosis \cite{Chaffer2025Incentivized}.







    
\section{Limitation and Future Work}
Our study has several limitations. First, the findings are constrained by a small sample size, primarily due to recruitment challenges. Developing and deploying DeAgents involves high technical barriers, which naturally limit the pool of potential stakeholders. In addition, the DeAgents community places a strong emphasis on personal privacy, and many individuals do not make their contact information publicly available, further complicating recruitment efforts. As a result, we primarily relied on snowball sampling through participants' personal networks. This approach may introduce bias, as it tends to include individuals with similar backgrounds or affiliations rather than providing a random or representative cross-section of the broader community. \added{In addition, DeAgents, as an emerging technology, is still in the early stages. The study focuses on the motivation and potential sociotechnical and political impact. Thus, the stakeholders involved in the study are mainly developers and industrial pioneers. We leave the work that engages more stakeholders for future work.}

Given the dilemmas we identified in the governance of DeAgents, we advocate that future work focus on exploring how to design systems that are both autonomous and accountable. This includes developing mechanisms that align agent behavior with human values over time, even as agents operate independently. Researchers should also investigate how to introduce forms of oversight or intervention that do not compromise decentralization, such as community-based safeguards or reversible controls. In addition, more work is needed to understand the legal, ethical, and social implications of delegating decision-making to non-human agents, particularly in high-stakes domains. Finally, future studies could explore hybrid governance models where humans and DeAgents collaborate, ensuring that AI-supported systems remain responsive, trustworthy, and aligned with the needs of their communities.

\section{Conclusion}
Decentralized AI agents promise a radical re-allocation of trust from human intermediaries to cryptographically verifiable code that exhibits properties of automaticity and potentially self-sovereignty. Yet in our qualitative study—the first to examine builders’, investors’, and researchers’ motivations and experiences with DeAgent launch, we find that this shift creates a “trustless-versus-trustworthy” paradox: once an autonomous agent is sealed into an immutable substrate, human deployers gain tamper-resistance but lose straightforward mechanisms for oversight, liability, and redress. To resolve this tension, future work must design revocable constitutional safeguards that preserve self-sovereignty, establish protocol-level identity and reputation layers (e.g., “Know-Your-Agent” registries) that allow agents to earn or lose privileges without central gatekeepers, and develop legal–HCI frameworks that render autonomous decisions explainable and accountable to human publics.

\bibliographystyle{ACM-Reference-Format}
\bibliography{reference}

\clearpage
\appendix

\section{Appendix}

\subsection{Interview Protocol}
\label{appendix:interview}
\textbf{Introduction and Consent}\\
Before starting the interview, participants will be informed of the research purpose and their right to withdraw at any time. The interview will be audio-recorded with the participant’s permission solely for note-taking and transcription purposes. Recordings will be kept confidential and stored securely. \\
\textbf{Background and Experience}
  \begin{itemize}
    \item Could you describe your background and experience with blockchain technology and decentralized AI?
    \item Have you deployed agents?
    \item What’s your role in the landscape of DeAgents?
  \end{itemize}
\textbf{Motivations and Expectations}
  \begin{itemize}
    \item What initially drew your interest in deploying Self-Sovereign AI Agents (DeAgents)?
    \item What specific benefits have you experienced or do you expect to experience from using decentralized AI? (e.g., initial model offering, team dynamics)
    \item How do you define “self-sovereignty” in the context of DeAgents, and why is it important to you?
    \item What role do trust and privacy play in your decision when deploying decentralized AI agents?
  \end{itemize}
\textbf{Challenges and Barriers}
  \begin{itemize}
    \item What specific technical or infrastructural challenges have you encountered when deploying DeAgents?
    \item How do you address or overcome these technical challenges? Are there any tools or solutions you rely on?
    \item Have you encountered organizational, legal, or community resistance to decentralized AI? If so, can you share examples of how you've navigated these challenges?
    \item Have the outcomes of your project aligned with your initial expectations? Why or why not?
  \end{itemize}
\textbf{Governance and Future Outlook}
  \begin{itemize}
   \item What if AI agents start acting against stakeholder interest (e.g., dumping tokens)? How do you think about the moral hazard involved?
    \item Have you considered the possibility that DeAgents might exceed your expectations and control? What does “ungovernability” mean to you in this context?
    \item How do you envision the future development of decentralized AI ecosystems?
    \item What governance structures do you think will be most effective for DeAgents moving forward?
    \item Do you think decentralized AI will face significant regulatory hurdles in the future? If so, what kinds of regulation might emerge?
    \item Given that self-sovereign DeAgents operate in TEEs, what if something goes wrong and you can’t stop it? Have you thought about integrating kill switches into the design? Why or why not?
  \end{itemize}
\textbf{Concluding Questions}
  \begin{itemize}
    \item What advice would you offer others before considering the adoption of DeAgents?
    \item If you were to do a version 2 of the project, what would you change?
    \item Are there any features or improvements you’d like to see in decentralized AI systems to enhance their usability or adoption?
    \item Is there anything else you'd like to share regarding decentralized AI or the deployment of DeAgents?
  \end{itemize}

\section{Disclosure of the usage of LLM}
We used ChatGPT (GPT5 model\cite{openai2024gpt}) to facilitate the writing of this manuscript. The usage includes:
\begin{itemize}
    \item Turn Excel format tables into LaTeX format tables
    \item Correct grammar mistakes and spelling
    \item Polish the existing writing by prompts like ``Find me a synonym of X,'' ``What is the noun/adjective form of X'' and ``Shorten this sentence without changing its content.''
\end{itemize}

\end{document}